\documentclass[bibyear]{aa} % for a printer version
\usepackage{graphicx}
\usepackage{txfonts}
\usepackage{natbib}
\bibpunct{(}{)}{;}{a}{}{,} % to follow the A&A style
\usepackage{amstext}

\begin{document} 

   \title{Chaotic enhancement of dark matter density in binary systems}

   \author{G.~Rollin\inst{1}
          \and
          J.~Lages\inst{1}
          \and
          D.~L.~Shepelyansky\inst{2}
          }

   \institute{Institut UTINAM, Observatoire des Sciences de l'Univers THETA,
              CNRS \& Universit\'e de Franche-Comt\'e, 25030 Besan\c{c}on, France\\
              \email{rollin@obs-besancon.fr}\\
              \email{jose.lages@utinam.cnrs.fr}
         \and
              Universit\'e de Toulouse, UPS,
              Laboratoire de Physique Th\'eorique (IRSAMC)
              F-31062 Toulouse, France;
              CNRS, LPT (IRSAMC), F-31062 Toulouse, France\\
              \email{dima@irsamc.ups-tlse.fr}
             }
             
   \date{Received ; accepted }
 
  \abstract{
We study the capture of 
galactic dark matter particles (DMP)
in two-body and few-body systems with a symplectic map description. 
This approach allows modeling the scattering of $10^{16}$ DMPs
after following the time evolution of the captured particle on about
$10^9$ orbital periods of the binary system.
We obtain the DMP density distribution inside such systems
and determine the enhancement factor of their density 
in a center vicinity compared to
its galactic value as a function of the mass ratio of the bodies 
and the ratio of the body velocity to the velocity of the galactic DMP wind. We 
find that the enhancement factor can be on the 
order of tens of thousands.}


   \maketitle
%
%________________________________________________________________

\section{Introduction}
In 1890, Henri Poincar\'e proved that the dynamics of the
three-body gravitational problem is generally non-integrable
\citep{poincare}. Even 125 years later, many aspects of this problem
remain unsolved. Thus the capture cross-section $\sigma$ of a particle that
scatters on the binary system of Sun and Jupiter
has only recently been determined, 
and it has been shown that $\sigma$ is much larger
than the area of the Jupiter orbit \citep{khriplovich2009,lages}. 
The capture mechanism is described by a symplectic dynamical
map that generates a chaotic dynamics of a particle.
The scattering, capture, and dynamics of a particle in a binary system
recently regained interest with the search for dark matter particles
(DMP) in the solar system and the Universe 
\citep{bertone2005,garrett,merritt}.
%%change
Thus it is important to analyze the capture and ejection mechanisms
of a DMP by a binary system. Such a system can be viewed
%
%as  a simplified galaxy model 
%represented by a binary system with a massive star 
as a binary system with a massive star
%in the center
and a light body orbiting it.
This can be the Sun and Jupiter, a star and a giant planet, or
a super massive black hole (SMBH) and a  light star or black hole (BH).
In this work we analyze the scattering process of 
DMP galactic flow, with a 
constant space density, in a binary system. One of the main questions here is
whether the density of captured DMPs in a binary system
can be enhanced compared to the DMP density of the scattering flow.

The results obtained by \cite{lages} show that
a volume density of captured DMPs at a distance of the 
Jupiter radius $r<r_p=r_J $ is enhanced by a factor
$\zeta \approx 4000$ compared to the
density of Galactic DMPs which are captured after one
one orbital period around the Sun and which have an
energy corresponding to velocities 
$v < v_{cap} \sim v_p \sqrt{m_p/M} \sim 1$km.s$^{-1} \ll u$.
Here, $m_p, M$ are the masses of the light  and massive bodies, respectively,
$u \approx 220$km.s$^{-1}$ is the average velocity of a Galactic DMP wind
for which, following \cite{bertone2005},
we assume a Maxwell velocity distribution:
$f(v) dv = \sqrt{54/\pi} v^2/u^3 \exp(-3v^2/2u^2) dv$.

%%change
Our results presented below show that for an SMBH binary system
with $v_{cap} > u$ there is a large enhancement
factor $\zeta_g \sim 10^4$ of 
%%change
the captured DMP volume density, taken at a distance 
of about a binary system size, compared to its galactic value
for all scattering energies (and not only for the 
DMP volume density at low velocities $v < v_{cap} \ll u,$
as discussed by \cite{lages}).
We note that the Galactic  DMP density is estimated at
$\rho_g \sim 4 \times 10^{-25}$g.cm$^{-3}$ , while
the  typical intergalactic DMP density
is estimated to be 
$\rho_{g0} \sim 2.5 \times 10^{-30}$g.cm$^{-3}$ \citep{garrett,merritt}.
At first glance, this high enhancement factor  $\zeta_g \sim 10^4$
seems to be rather unexpected because it apparently contradicts Liouville's theorem,
according to which the phase space density 
is conserved during a Hamiltonian evolution.
Because of this, it is often assumed \citep{gould,edsjo1}
that the volume (or space) DMP density cannot be enhanced
for DMPs captured by a binary system, and thus $\zeta_g \sim 1$.
Below we show that this restriction is not valid for the following
reasons: first, we have an open system where DMPs
can escape to infinity, being ejected from the binary
system by a time-dependent force induced by binary rotation.
This means that the dynamics is not completely Hamiltonian.
Second, DMPs are captured (or they linger, or are trapped) and 
are accumulated from continuum at negative coupled energies  near the binary
during a certain capture lifetime (although not forever).
Thus, the longer the capture lifetime, the higher the accumulated density.
Third, we obtain the enhancement for the volume density 
and not for the density 
in the phase space, for which the 
enhancement is indeed restricted by Liouville's  theorem. 
We discuss the details of this enhancement effect in the next sections.

The scattering and capture process of a DMP
in a binary system can be an important element of galaxy formation.
This process can also be 
useful to analyze cosmic dust and DMP interaction
with a supermassive black hole binary. This is expected to play a prominent role in galaxy formation, see
\cite{nature2015}.
Thus we hope that analyzing this process
will be useful for understanding the properties of velocity curves in galaxies, which was
started by \cite{zwicky} and \cite{rubin}.
We note that the velocity curves of captured DMPs in our binary system
have certain similarities with those found in real galaxies.

\section{Symplectic map description}
Following the approach developed by
\cite{petrosky}, \cite{halley}, \cite{tremaine1999}, and \cite{lages},
we used a symplectic dark map description
%%change
of the DMP dynamics in one orbital period of a DMP in a binary system
\begin{equation}
\label{eq1}
w_{n+1}=w_n+F(x_n) \; , \;\; x_{n+1}=x_n+w^{-3/2}_{n+1} \; ,
\end{equation}
where $x_n= t_n/T_p \; (mod \; 1)$ is given by time $t_n$ 
taken at the moment of DMP $n-th$ passage through perihelion,
$T_p$ is the planet period, and $w=-2E/m_d v_p^2$. Here
$E, m_d,   \text{and } v_p$ are the
energy, mass of the DMP, and the velocity of the planet or star.
The amplitude $J$  of the kick $F$-function is
proportional to the mass ratio $J \sim m_p/M$. The shape of $F(x)$
depends on the DMP perihelion distance $q$,
the inclination angle $\theta$ between the planetary plane $(x,y)$
and DMP plane, and the perihelion orientation angle $\varphi$
, as discussed by \cite{lages}.
In the following 
we use for convenience units with 
$m_d=v_p=r_p=1$ (here $m_d$ is the DMP mass, which 
does not affect the DMP dynamics in gravitational systems).
%%%%%%%%%%%%%%%%%%%%%%%%%%%%%%%%%%%%%%%%%%%%%%%%%%%%%%%%%

\begin{figure}
\resizebox{\hsize}{!}{\includegraphics{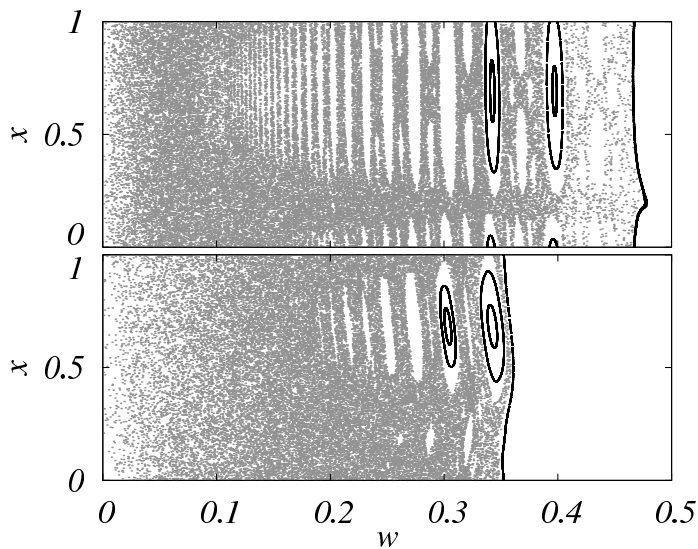}}
\caption{Poincar\'e sections for the dark map (\ref{eq1}) (top)
and the Kepler map  (\ref{eq2}) (bottom) 
for parameters of the Halley comet case in Eq. (\ref{eq1})
and $J=0.007$ in Eq. (\ref{eq2}) (see text).}
\label{fig1}
\end{figure}

%%%%%%%%%%%%%%%%%%%%%%%%%%%%%%%%%%%%%%%%%%%%%%%%%%%%%%%%%

For $q>r_p$ the amplitude $J$
drops exponentially with $q$ and  $F(x)= J \sin (2 \pi x),$
as shown by \cite{petrosky}.
This functional form of $F(x)$ is significantly
simpler than the real one at $q < r_p$ , while it still
produces chaotic dynamics at $0< w \ll 1$ and integrable motion
with invariant curves above a chaos border  
$w > w_{ch}$.
In this regime the map takes the form
\begin{equation}
\label{eq2}
w_{n+1}=w_n+J\sin (2\pi x_n) \; , \;\; x_{n+1}=x_n+w^{-3/2}_{n+1} \; .
\end{equation} 
 The same map describes a microwave ionization
of excited hydrogen atoms that is called the Kepler map,
see \citep[see][]{kepler,rydberg}. There, the Coulomb attraction
plays the role of gravity, while a circular planet rotation is
effectively created by the microwave  polarization.
The microwave ionization experiments
performed by \cite{koch} were made for three-dimensional
atoms, but the ionization process is still well
described by the Kepler map \citep[see][]{kochtheor,rydberg}.
These results provide additional arguments in favor of 
a simplified Kepler map description of DMP dynamics
in binary systems.  The dynamics of the Kepler map can be 
locally described by the Chirikov standard map
\citep[see][]{chirikov1979}.
%%change
We note that the approach based on the Kepler
map has recently been used to determine chaotic
zones in gravitating binaries, see \cite{shevchenko15}.

The similarity of dynamics of dark (\ref{eq1})
and Kepler  (\ref{eq2}) maps is also well visible from comparing 
their Poincar\'e sections, shown in Fig.~\ref{fig1},
for the typical dark map parameters corresponding to the Halley comet 
\citep[see Fig.1a in][]{lages} and the corresponding parameter $J$
of the Kepler map. 

To take into account that
$J$ decreases with $q,$ we use the relation
$J=J_0=const$ for $q<q_b$ and $J=J_0 \exp(-\alpha(q-q_b)))$
for $q \ge q_b$ (below $J$ is used instead of $J_0$). 
We use $ q_b=1.5$ and $\alpha=2.5$,
corresponding to typical dark map parameters
\citep[see Fig.1 in ][]{lages}, but we checked that
the obtained enhancement is not affected
by a moderate variation of $q_b \text{ or } \alpha$.
The simplicity of map (\ref{eq2}) allows increasing the number 
$N_p$ of injected DMPs by  a factor one hundred
compared to map (\ref{eq1}). 
The correspondence between  (\ref{eq1}) 
and (\ref{eq2}) is established by the
relation $J=5 m_p/M,$ which works
approximately for the typical parameters 
of Halley comet case. 

Of course, as discussed by \cite{lages},
the dark map and moreover the Kepler map
give an approximate description of 
DMP dynamics in binary systems. However, this
approach is much more efficient
than the exact solution of Newton equations
used by \cite{peter2009}, \cite{peter1}, and \cite{edsjo}
and allows obtaining results with
very many DMPs injected
during the lifetime of the solar system (SS)
$t_S=4.5 \times 10^9$ years. The validity
of such a map description 
is justified by the results obtained by
\cite{petrosky}, \cite{halley}, \cite{tremaine1999}, \cite{lages}, \cite{rollin15a}, and \cite{kochtheor}.

\section{Capture cross-section}
%%%%%%%%%%%%%%%%%%%%%%%%%%%%%%%%%%%%%%%%%%%%%%%%%%%%%%%%%

\begin{figure}
\resizebox{\hsize}{!}{\includegraphics{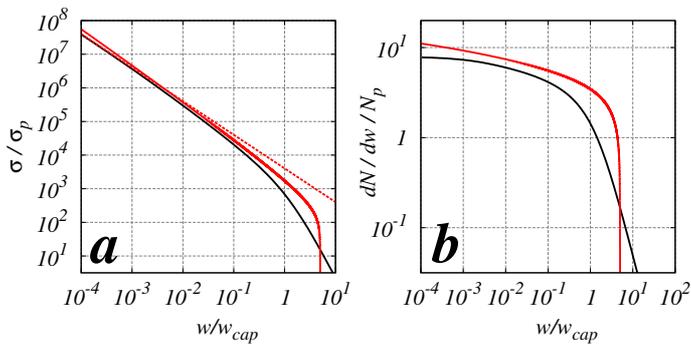}}
\caption{{\it (a)} Dependence of the capture
cross-section $\sigma$ on DMP energy $w$
for Sun-Jupiter (black curve, data from Ref.[8])
and 
for the Kepler map at $J=0.005$ (red curve);
the dashed line shows the dependence $\sigma \propto 1/|w|$.
{\it (b)} Dependence of the rescaled 
captured number of DMPs on energy $w$ for 
the models of the left panel. Here $w_{cap}=0.001$.}
\label{fig2}
\end{figure}

%%%%%%%%%%%%%%%%%%%%%%%%%%%%%%%%%%%%%%%%%%%%%%%%%%%%%%%%%

The capture cross-section $\sigma$
is computed as previously described by \cite{lages}
with $\sigma(w)/\sigma_p = (\pi^2 r_p |w|)^{-1} 
\int_0^{2\pi} d \theta \int_0^\pi d\varphi \int_0^{\infty} dq h(q,\theta,\varphi)$,
where $h$ is a fraction of DMPs captured
after one map iteration from $w<0$ to $w>0$, 
given by an interval length  inside the $F(x)$ envelope 
at $|w|=const$, $\sigma_p=\pi r_p^2$. 
The equation for $\sigma(w)$ is based on the
expression for the scattering impact parameter
$ r_d^2=2qr_p/|w|$.
For the Kepler map the 
$h-$function only depends on $q,$ and the numerical 
computation is straightforward. 
The  differential energy distribution of captured DMPs 
is $dN/dw =\sigma(w) n_g f(w)/2$ with $n_g=\rho_g/m_d$.

The results for 
$\sigma(\omega)$ and $dN/dw/N_p$, obtained for
maps (\ref{eq1}) and (\ref{eq2}), are shown in Fig.~\ref{fig2}.
Here $N_p=\int_0^1 dw n_g \sigma_p v_p^2 f(w)/2$ is 
the  number of DMPs crossing the planet orbit area per unit of time.
The results of Fig.~\ref{fig2} show that both maps 
give similar results, which provides additional support
for the Kepler map description. 
The theoretical dependence $\sigma \propto 1/|w|$,
predicted by \cite{khriplovich2009},
is clearly confirmed.
The only difference between maps (\ref{eq1}) and (\ref{eq2}) is
that the kick amplitude $J \approx 5m_p/M$ for (\ref{eq2}) is restricted,
and thus after one kick we may have only $|w| \leq J$, while
for (\ref{eq1}) some orbits can be captured with
$|w| > J =5m_p/M$ as a result of close encounters. 
However, the probability of such events is low.

\section{Chaotic dynamics}

The injection, capture, evolution, and escape of DMPs is 
computed as described by \cite{lages}:
we numerically modeled a constant flow of scattered DMPs
with an energy distribution  $d N_s  = \sigma(w) v_p^2f(w) dw/2$ 
per time unit (we used $q \leq q_{max}=4r_p$). 
For Jupiter we have $u \approx 17 \gg 1$
and  $d N_s \propto  dq d w$. However, for an SMBH we can 
have $u^2 < J$ so that one kick captures
almost all the DMPs from the galactic distribution $f(w)$.
In this case, we used the whole distribution $f(w)$
($w=v^2$).
Map (\ref{eq2}) is simpler than (\ref{eq1})
since the kick function only depends on $q,$
which allows performing simulations with more DMPs.

%%%%%%%%%%%%%%%%%%%%%%%%%%%%%%%%%%%%%%%%%%%%%%%%%%%%%%%%%

\begin{figure}
\resizebox{\hsize}{!}{\includegraphics{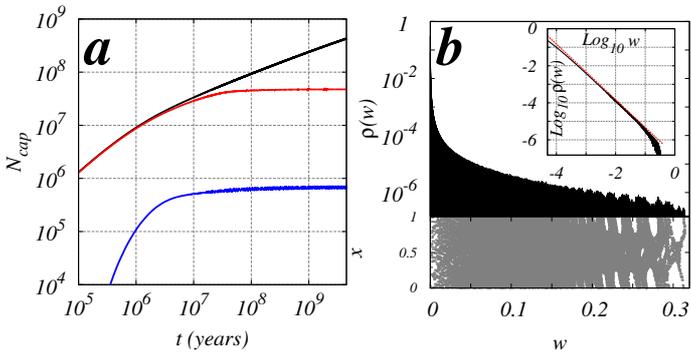}}
\caption{{\it (a)} Number $N_{cap}$  
of captured DMPs as a function of time $t$ in years
for  the energy range 
$w >0$ (black curve), $w >4 \cdot 10^{-5}$ corresponding to
half the distance between Sun and 
the Alpha Centauri system (red curve), $w >1/20$
corresponding to $r<100 {\rm AU}$
(blue curve); $N_J=4 \times 10^{11}$ DMPs are injected
during SS lifetime $t_S$; data are obtained from 
the map (\ref{eq2}) at $J=0.005$, $u=17$
corresponding to the Sun-Jupiter case.
{\it (b)} The top part shows the density distribution 
$\rho(w) \propto dN/dw$ in energy
at time $t_S$
(normalized as $\int_0^\infty \rho dw=1$),
the bottom part shows the Poincar\'e section 
of the map (\ref{eq2}); the inset shows 
the density distribution of the captured DMPs in $w$
(black curve), the red line shows the slope -3/2.}
\label{fig3}
\end{figure}

%%%%%%%%%%%%%%%%%%%%%%%%%%%%%%%%%%%%%%%%%%%%%%%%%%%%%%%%%

The scattering and evolution processes were followed during the whole 
lifetime $t_S$ of the SS.
The total number of DMPs, injected during time $t_{S}$
for $|w| \leq J$ and all $q$ is $N_J$. For the Kepler
map the highest value is $N_J = 4 \times 10^{11}$ 
, which is $100$ times higher than  for the dark map.

%%%%%%%%%%%%%%%%%%%%%%%%%%%%%%%%%%%%%%%%%%%%%%%%%%%%%%%%%

\begin{figure}
\resizebox{\hsize}{!}{\includegraphics{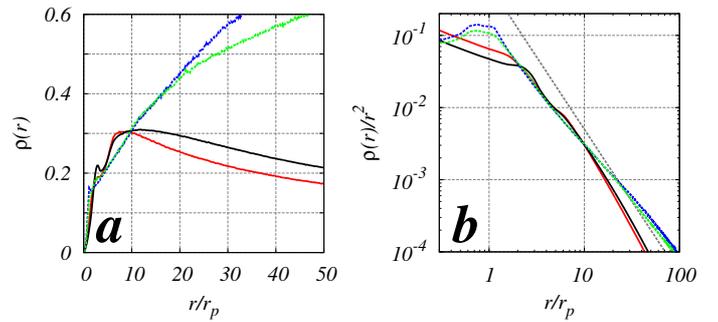}}
\caption{{\it (a)} Stationary radial density $\rho(r) \propto dN/dr$
from the Kepler map at $J=0.005$
with  $u=17$ at time $t_S$ (red curve) and 
$u=0.035$ at time $t_u \approx 4 \times 10^8 T_p$(black curve);
data from the dark map 
at $m_p/M=10^{-3}$ are shown by
the blue curve at $u=17$ and time $t_S$ for the Sun-Jupiter case,
and by the green curve at $u=0.035$ and $t_S$  for the SMBH;
the normalization is fixed as $\int_0^{6r_p} \rho dr =1$, $r_p=1$.
{\it (b)} Volume density $\rho_v=\rho/r^2$ from 
the data of panel {\it (a)},
the dashed line shows the slope $-2$.}
\label{fig4}
\end{figure}

%%%%%%%%%%%%%%%%%%%%%%%%%%%%%%%%%%%%%%%%%%%%%%%%%%%%%%%%%

The time dependence $N_{cap}(t)$ for the Kepler map, 
shown in Fig.~\ref{fig3}, is
very similar to that found for the dark map
by \cite{lages}. For a finite SS region $w> 1/20$
the growth of $N_{cap}(t)$ saturates
after a time scale of $t_d \approx 10^7$ years. 
This scale approximately corresponds to a diffusive escape time
$t_d \sim 12\text{ }   {\rm years} /D \sim  10^6 {\rm \text{
}years,}$ where
the diffusion rate is taken in a random phase approximation to
be  $D \approx J^2/2$ \citep[see, e.g.,][]{kepler}. The diffusive spreading
extends from $w \sim 0$ up to chaos border $w_{ch} \approx 0.3$.
This value  agrees well with the
theoretical value $w_{ch}=(3\pi J)^{2/5} = 0.29$ obtained 
from the Chirikov criterion \citep{chirikov1979} 
\citep[see discussion for DMP dynamics in][]{petrosky,kepler,khriplovich2009}.
The validity of the Chirikov criterion
in this system was
also demonstrated in \cite{shevchenko15}.
As for the dark map, we obtain a density distribution of
$\rho(w) \propto 1/w^{3/2}$ , corresponding
to the ergodic estimate according to which
$\rho(w)$ is proportional to time period at a given $w$.
The results of Figs.~\ref{fig1},\ref{fig2}, and \ref{fig3} 
confirm the close similarity of dynamics 
described by maps (\ref{eq1}) and (\ref{eq2}).

\section{Radial variation of the dark matter density}

To compute the DMP density, we considered captured orbits $N_{AC}$ with
$w>4 \times 10^{-5}$. The radial density 
$\rho(r)$ was computed by the method described by \cite{lages}:
$N_{AC}$ were determined at instant time 
$t_S$; for them the dynamics in real space was recomputed 
during a time period  $\Delta t \sim 100 $ {\rm years} of planet.
The value of $\rho(r)$ was computed by averaging over $k=10^3$
points randomly distributed over $\Delta t$ for all $ N_{AC}$
orbits.

%***************************************************************************************************************************************************%
We also checked that a semi-analytical averaging, using an exact density distribution over Kepler ellipses for each of $N_{AC}$ orbits, gives the same 
result: assuming ergodicity $\rho_{w,q}(r)dr=w^{3/2}dt/2\pi$ and using Kepler's equation, the radial density of the DMPs on a given orbit is
$\rho_{w,q}(r)=\left(rw^2/2\pi\right)\left(\left(1-qw\right)^2-\left(1-rw\right)^{2}\right)^{-1/2}$, then adding the radial density of each
$N_{AC}$ orbit, we retrieve the DMP radial density $\rho(r)$ shown in Fig. \ref{fig4}.
%***************************************************************************************************************************************************%
From the obtained space distribution
we determine a fraction $\eta_{r_i}$ of $N_{AC}$ DMP orbits
located inside a range
$0 \leq r \leq r_i$ by computing
$\eta_{r_i} = \Delta N_i/(k  N_{AC}),$
where $\Delta N_i$ is the number of points inside the above range
(we used $r_i/r_p=0.2, 1, \text{and}   \text{ }6$). 

In Fig.~\ref{fig4} we show the dependence of radial $\rho(r)$
and volume $\rho_v=\rho/r^2$ densities on distance $r$.
For the Kepler map data, the density $\rho(r)$ has a 
characteristic maximum at $r_{max}$ that
is  determined by the chaos border position 
$r_{max} \approx 2/w_{ch}$ 
(this dependence, as well as the relation $w_{ch} = (3\pi J)^{2/5}$,
is numerically confirmed for the 
studied range $10^{-3} < J < 10^{-2}$ for the Kepler map
with a given fixed $J$).
The density profile $\rho(r)$ is not 
sensitive to the value of $u$ 
and remains practically  unchanged for
$u=17, \; 0.035$.
For the dark map a variation of the kick function
with $q$ and angles leads to a variation of $w_{ch}$
that leads to a slow growth of $\rho$ at large $r$.
A power-law  fit of $\rho_v \propto 1/r^\beta$
in a range $2 < r < 100$ gives
$\beta  \approx 2.25 \pm 0.003$ for the Kepler map data
and $\beta = 1.52 \pm 0.002$ for the dark map.
We attribute the difference in $\beta$ values to
a larger fraction of integrable islands
for the dark map, as is visible in Fig.~\ref{fig1}
for typical parameters.
We note that an effective range of radial variation 
is bounded by the kick amplitude with $r < r_{cap} \approx 1/J,$
and in the range $r_p<r<r_{cap}$ the data are compatible with
$\rho \sim const$ (dashed line in Fig.~\ref{fig4}b).

\begin{figure}
\resizebox{\hsize}{!}{\includegraphics{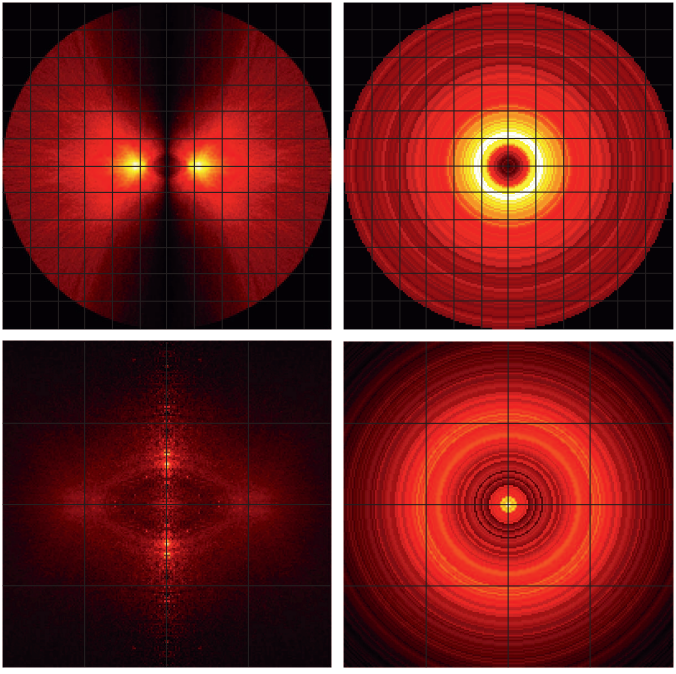}}
\caption{Density of captured DMPs at present time $t_S/T_p \approx 4 \times 10^8$
for the dark map at $m_p/M=10^{-3}$ and $u/v_p=0.035$
{\em Top panels:} DMP surface density 
$\rho_s \propto dN/dzdr_\rho$ 
shown at the {\em left}  
in the cross plane $(0,y,z)$ perpendicular to the planetary orbit
(data are averaged over $r_\rho=\sqrt{x^2+y^2}=const$),
at the {\em right} in the planet plane $(x,y,0)$;
only the range $|r| \leq 6$ around the center is shown. 
{\em Bottom panels:} corresponding 
DMP volume density $\rho_v \propto dN/dxdydz$ at the {\em left}
in the plane  $(0,y,z)$, at the {\em right} in the planet plane 
$(x,y,0)$;
only the range $|r| \leq 2$ around the SMBH is shown. 
The color is proportional to the density
with yellow/black for maximum/zero density.}
\label{fig5}
\end{figure}

We note that the value of $u$ does not significantly
affect the density variation with $r$ , as is clearly seen
in Fig.~\ref{fig4}. The spacial density
distribution of computed from the dark map at $u=0.035$
shown in Fig.~\ref{fig5} is also very similar to those
at $u=17$ \citep[see Fig.5 by][]{lages}.
This independence of $u$
arises because $\rho(r)$
is determined by the dynamics at $w>0,$
which is practically insensitive to the DMP energies 
at $-J<w<0$  that are captured by one kick.

%%%%%%%%%%%%%%%%%%%%%%%%%%%%%%%%%%%%%%%%%%%%%%%%%%%%%%%%%

%%%%%%%%%%%%%%%%%%%%%%%%%%%%%%%%%%%%%%%%%%%%%%%%%%%%%%%%%

\section{Enhancement of dark matter density}

To determine the enhancement of the DMP density
captured by a binary system we followed
the method developed by \cite{lages}.
We computed the total mass of DMP flow 
crossing the range $q \leq 4 r_p$ during time $t_S$:
$M_{tot}= \int_0^{\infty} dv v f(v) \sigma \rho_g t_S 
\approx  35 \rho_g t_S k r_p M/u, $
where we used the cross-section $\sigma = \pi r_d^2 = 8\pi k M r_p/v^2$
for injected orbits with $q \leq 4 r_p$, $w=v^2$,
$k$ is the gravitational constant.
For SS at $u/v_p \approx 17$ we have $M_{tot} \approx 0.5 \cdot 10^{-6} M$. 

From the numerically known fractions
$\eta_{ri}$ of the previous section 
and the fraction of captured orbits 
$\eta_{AC}=N_{AC}/N_{tot}$ we
find the mass $M_{ri}=\eta_{ri} \eta_{AC} M_{tot}$
inside the volume $V_i= 4\pi r_i^3/3$
of radius $r<r_i$ ($r_i=0.2 r_p; r_p; 6 r_p$). Here 
$N_{tot}$ is the total number of injected orbits
during the time $t_S$ , while
the number of orbits injected in the range
$|w|<J$ (only those can be captured)
is $N_J = N_{tot} ( \int_0^{J} dw f(w)/w)/( \int_0^{\infty} dw f(w)/w)$.
For $J \ll u^2$ we have 
$\kappa= N_{tot}/N_J = 2u^2/(3J) \approx 3.8 \times 10^4$
for $u/v_p =17$ and $\kappa=1$ for $u/v_p=0.035$ at $J=0.005$.
Thus for $u/v_p=17$ the number of  orbits, injected 
at $0 <|w|<J$, $N_J = 4 \times 10^{11}$ 
, corresponds to the total
number of injected orbits $N_{tot} \approx 1.5 \times 10^{16}$.
Finally, we obtain the global density enhancement factor 
$\zeta_g(r_i)=\rho_v(r_i)/\rho_g \approx 
16 \pi \eta_{ri} \eta_{AC} (r_p/r_i)^3 \tau_S v_p/u$,
where $\tau_S=t_S/T_p$ is the injection time expressed
in the number of planet periods  $T_p=2\pi r_p/v_p$.
For $u^2 \gg J$ it is useful to determine 
the enhancement $\zeta = \rho_v(r_i)/\rho_{gJ}$ 
of the scattered galactic density in the range 
$0<|w|<J,$ whose density is $\rho_{gJ} \approx 1.38 \rho_g J^{3/2} (v_p/u)^3$.
Thus $\zeta = 0.72  \zeta_g (u/v_p)^3/J^{3/2}$.

\begin{figure}
\resizebox{\hsize}{!}{\includegraphics{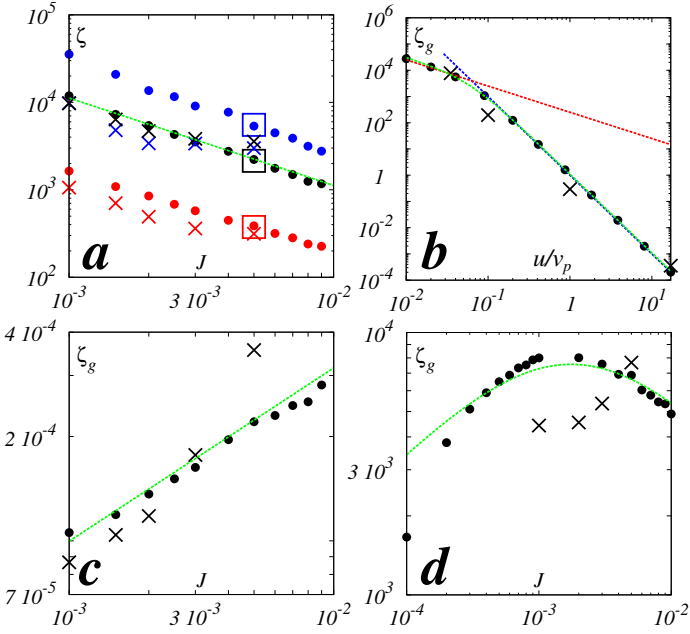}}
\caption{Dependence of the DMP density enhancement
factor $\zeta=\rho_v(r_i)/\rho_{gJ}$
on $J$ at $u/v_p=17$ (Jupiter); here $\rho_{gJ}$
is the galactic DMP volume density
for an energy range of $0<|w|<J$ and 
$r_i    =0.2 r_p, r_p, 6 r_p$ (blue, black, red);
points and squares show results for map (\ref{eq2})
with the number of injected particles 
$N_J= 4 \times 10^{9}$ and $4 \times 10^{11}$
, respectively; crosses show data for map (\ref{eq1})
with $N_J= 4 \times 10^{9}$ and $J=5m_p/M$.
{\it (b)}  Dependence of the galactic enhancement factor
 $\zeta_g=\rho_v(r_i    )/\rho_{g}$
on $u/v_p$ at $r_\zeta=r_p$ and $J=0.005$ in (\ref{eq2}) (points)
and $m_p/M=0.001$ in  (\ref{eq1}) (crosses),
here $\rho_{g}$ is the global galactic density;
lines show dependencies $\zeta_g \propto 1/u$ (red)
and $\zeta_g \propto 1/u^3$ (blue).
 {\it (c)}  Dependence of $\zeta_g$ on $J$ at $u/v_p=17$; 
 {\it (d)} the same at  $u/v_p=0.035$,
parameters of symbols are as in {\it (a),(b)}.
The green curve shows theory (\ref{eq3}) in all panels. }
\label{fig6}
\end{figure}

The results of the DMP density enhancement factors
$\zeta$ and $\zeta_g$ are shown in Fig.~\ref{fig6}.
At $(u/v_p)^2 \gg J$ we have $\zeta \gg 1$ and $\zeta_g \ll 1$.
At $u/v_p=17$ we find that $\zeta \propto 1/J$
(the fit gives exponent $a=1.04 \pm 0.01$)
and $\zeta_g \propto \sqrt{J}$ 
(the fit exponent is $a=0.46 \pm 0.1$)
in agreement with the above relation
between $\zeta$ and $\zeta_g$. In general,
we have $\zeta_g \propto 1/u$ for $u/v_p \ll \sqrt{J}$
and $ \zeta_g  \propto 1/u^3$ for $u/v_p \gg \sqrt{J}$.
There is only weak variation of 
$\zeta_g$ with $J$ for $u/v_p \ll \sqrt{J}$.
The values of $\zeta$ and $\zeta_g$
have similar values for the dark and Kepler maps
(a part of the fact that at $r_i=0.2 r_p$ and $r_i=r_p$
the dark map has approximately the same $\zeta$
since there $\rho_v(r) \sim const $ for $r \leq r_p$).   

%%%%%%%%%%%%%%%%%%%%%%%%%%%%%%%%%%%%%%%%%%%%%%%%%%%%%%%%%

%%%%%%%%%%%%%%%%%%%%%%%%%%%%%%%%%%%%%%%%%%%%%%%%%%%%%%%%%

\begin{figure}
\resizebox{\hsize}{!}{\includegraphics{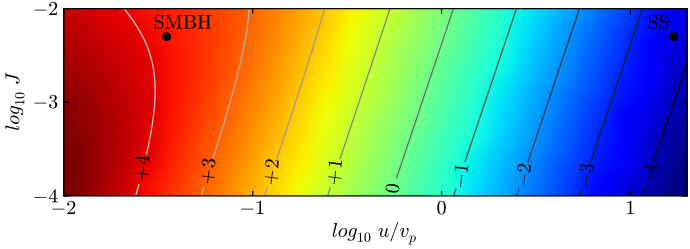}}
\caption{Logarithm of DMP density enhancement factor $\log_{10}\zeta_g$ 
from (\ref{eq3}), shown by color and $\log$ value-levels, 
as a function of $u/v_p$ and
$J$; two points are for $J=0.005$,  $u/v_p=17$ (SS)
and    $u/v_p=0.035$ (SMBH; such $v_p$ is about $2 \%$ of the
light velocity)).}
\label{fig7}
\end{figure}

All these results can be summarized by the following
formula for the chaotic enhancement factor of DMP density
in a binary system:
\begin{equation}
\label{eq3}
\zeta_g = A \sqrt{J} (v_p/u)^3/[1+B J (v_p/u)^2] \; ,  
J=5 m_p/M \; .
\end{equation}
Here  $\zeta_g$ is given for DMP density at $r_i=r_p$ and $A \approx 15.5$,
$B \approx 0.7$.
This formula describes the numerical data of Fig.~\ref{fig6}
well.
For $(u/v_p)^2 \gg J$ we have $\zeta_g \ll 1,$ but we still  have
an enhancement of  $\zeta = 0.72 \zeta_g (u/v_p)^3/J^{3/2}  \approx  0.72A/J \gg 1$. 
For $(u/v_p)^2 \ll J$ we have the global enhancement
$\zeta_g \approx 22(v_p/u)/\sqrt{J} \gg 1$.
The color representation of dependence (\ref{eq3}) is shown in Fig.~\ref{fig7}.

%%%%%%%%%%%%%%%%%%%%%%%%%%%%%%%%%%%%%%%%%%%%%%%%%%%%%%%%%

%%%%%%%%%%%%%%%%%%%%%%%%%%%%%%%%%%%%%%%%%%%%%%%%%%%%%%%%%

Equation (\ref{eq3}) can be understood on the basis of 
simple estimates. The total captured mass $M_{cap} \approx M_{AC}$ 
is  accumulated during the diffusive time $t_d$ and hence
$M_{cap} \sim v_p^2 J t_d M_{tot}/(\pi u^2 t_S) \sim 
\rho_g \tau_d J (v_p/u)^3$ , where $\tau_d=t_d/T_p$ 
, and we omit  numerical coefficients.
This mass is concentrated inside a radius $r_{cap} \sim 1/J$
so that at  $r \sim 1/J$ the volume density
is $\rho_v(r=1/J) \sim M_{cap}/r_{cap}^3 \sim \rho_g J^2 w_{ch}^2 (v_p/u)^3 \sim
\rho_{gJ} J^{1/2} w_{ch}^2 \sim \rho_{gJ} J^{1.3} $, where we use a relation
$\tau_d \sim w_{ch}^2/J^2 \sim 1/J^{6/5}$.
(Our modeling of the injection
process in the Kepler map with a constant injection flow
in time, counted as the number of map iterations, shows that the number of absorbed particles
scales as $N_K \sim \tau_d \sim J^{-6/5}$ at small $J$.)
It is important to stress that $\rho_v(r=1/J) \ll \rho_{gJ}$
in contrast to the naive expectation that $\rho_v(r=1/J) \sim \rho_{gJ}$.
Using our empirical density decay $\rho_v \propto 1/r^\beta$
with $\beta \approx 2.25$ for the Kepler map, we obtain 
$\zeta \propto 1/J^{0.95}$ , which is close to 
the dependence $\zeta \sim 1/J$ and $\zeta_g \sim J^{1/2}/(u/v_p)^3$ 
from (\ref{eq3}) at $u^2 \gg J$. 
For the dark map we have $\beta \approx 1.5$ 
but $w_{ch} \sim const$ as a result of the 
sharp variation of $F(x)$ with $x$ , which again gives $\zeta \sim 1/J$.
It is difficult to obtain the exact analytical derivation
of the relation $\zeta \sim 1/J$ due to contributions of
different $q$ values (which have different 
$\tau_d$) and different kick shapes in (\ref{eq1})
that affect $\tau_d$ and 
the structure of chaotic component.
In the regime $(u/v_p)^2 \ll J$ the entire energy range of
the scattering flow is absorbed by one kick,
and $M_{cap}$ is increased by a factor $(u/v_p)^2/J,$
leading to an increase of $\zeta_g$ by the same
factor, which yields $\zeta_g \propto v_p/(u \sqrt{J}),$
in agreement with (\ref{eq3}).

We note that for galaxies the value of exponent $\beta$
is debated \citep[see][]{merritt}.
For the adiabatic growth model, we have 
$2.25 \leq \beta \leq 2.5, $  which is close to the value
obtained from our symplectic map simulations. 

%%%%%%%%%%%%%%%%%%%%%%%%%%%%%%%%%%%%%%%%%%%%%%%%%%%%%%%%%

\begin{figure}
\resizebox{\hsize}{!}{\includegraphics{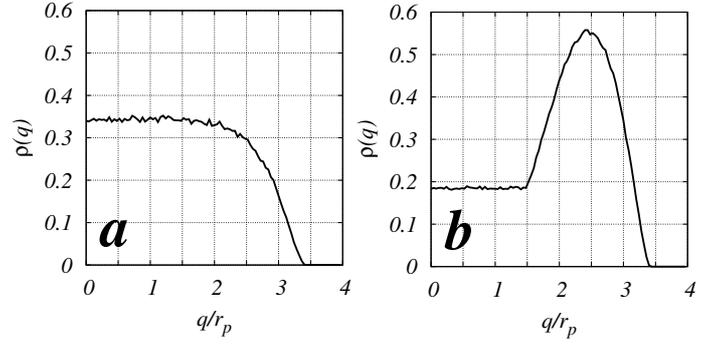}}
\caption{Density distribution of DMPs  $\rho(q)$ over $q$
obtained from the Kepler map at $J=0.005$ and  
 time $t_u \approx 4 \times 10^8 T_p$:
{\it (a)}  $u/v_p=17$;
{\it (b)}  $u/v_p=0.04$;
the density is normalized to unity
($\int_0^{\infty} \rho dq/r_p=1$).}
\label{fig8}
\end{figure}

%%%%%%%%%%%%%%%%%%%%%%%%%%%%%%%%%%%%%%%%%%%%%%%%%%%%%%%%%

The nontrivial properties of the distribution of the captured DMPs
in $q$ are shown in Fig.~\ref{fig8} in a stationary regime
at times $t_S/T_p \approx 4 \times 10^8$ for the Kepler map.
While for $u/v_p \sim 17 \gg 1$ we have a smooth drop
of DMP density $\rho(q)$ at $q>1.5r_p$ 
, for $u/v_p =0.04 \ll 1$
we have an increase of  $\rho(q)$ by a factor $3$
for $q/r_p \approx 2.5$
compared to $q/r_p \approx 1$.
We attribute this variation to different
capture conditions at $u \gg \sqrt{J} v_p$, where only 
DMPs at low velocities are captured by one kick,
and $u \ll \sqrt{J} v_p$, where practically all DMPs are captured
by one kick. As a result of the dependence of $J$ on $q,$
we also have various diffusive timescales $t_d \propto 1/J^2$
that can affect the contribution of
the DMPs at different $q$ values in the volume density distribution
on $r$.

Finally, we stress the importance of the obtained result 
of large enhancement factors $\zeta$ and $\zeta_g$. 
This result is drastically different from the frequent claims
that there is no enhancement of the DMP density in the center
vicinity of a binary system 
compared to its galactic value because of 
 the Liouville theorem, which implies
that the density of DM in the phase space 
is conserved during the evolution \citep{gould,edsjo1}.
However, this statement does not take into account the
actual dynamics of captured DMPs.
Indeed, the galactic space density $\rho_g$ is obtained from all energies
of DMPs in the Maxwell distribution.
The analysis of symplectic DMP dynamics
shows that DMPs at large $q \gg 1$ are not
captured, while DMPs with $q \sim 1$
are captured, and by diffusion, they penetrate up to
high values $w \sim w_{ch}$ , thus accumulating
DMPs with typical distance values $r \sim 1/w_{ch}$.
The symplectic map approach also determines an effective size
of our binary system of $r_{cap} \sim 1/J$
corresponding to an energy range $w \sim J$.
If we assume that the DMP density in this range is
the same as its galactic value, then we should conclude
that the enhancement factor should be 
$\zeta_g \sim (r_{cap}/r_p)^\beta \sim 1/{w_{cap}}^{\beta} \sim
1/J^{\beta} \sim 1.5 \times 10^5$
for typical values $J=0.005$ and $\beta=2.25$
(we consider here the case $u/v_p \ll \sqrt{J}$). This estimate
gives a value $\zeta_g$ 
that is even higher than that given by relation (\ref{eq3}).
In fact, relation (\ref{eq3}) takes into account that
only bounded values of $q$ are captured, it also estimates the chaos region, where  DMPs are accumulated 
during the chaotic diffusion process, populating a part of
the phase space volume from $w \sim 0$ up to
$w \sim w_{ch} \sim 1$. This gives a lower value of $\zeta_g$
than the above simplified estimate. 
 We also note that at $u/v_p \ll \sqrt{J}$
the typical kinetic energy of an ejected DMP $J v_p^2$ 
is significantly higher than
the typical DMP energy $u^2$ in the galactic wind.
For these reasons, there is no contradiction with the Liouville
theorem, and a large enhancement of the captured DMP density is possible.

\section{Few-body model}

%%%%%%%%%%%%%%%%%%%%%%%%%%%%%%%%%%%%%%%%%%%%%%%%%%%%%%%%%

\begin{figure}
\resizebox{\hsize}{!}{\includegraphics{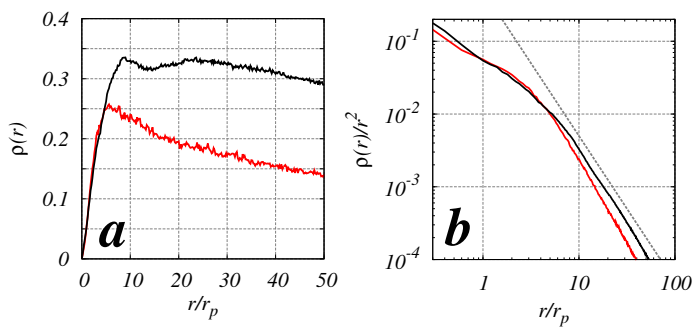}}
\caption{{\it (a)} Radial density $\rho(r) \propto dN/dr$
 for the Kepler models of SS
(red curve) and  SMBH binary (black curve)   at
$t_S/T_p \approx 4 \times 10^8$;
the normalization is fixed as $\int_0^{6r_p} \rho dr =1$,
$r_p=1$ for the fifth body.
{\it (b)} Volume density 
$\rho_v=\rho/r^2$ from the data of {\it (a)},
the dashed line shows the slope $-2$ (see text for details).}
\label{fig9}
\end{figure}

%%%%%%%%%%%%%%%%%%%%%%%%%%%%%%%%%%%%%%%%%%%%%%%%%%%%%%%%%
%%change
Above we considered the DMP capture in a two-body gravitating system.
We expect that a central SMBH binary dominating 
the galaxy potential can be viewed as a simplified 
galaxy model. Recent observations of \cite{nature2015}
indicate that such systems may exist.
Within the Kepler map approach it is easy to analyze
the whole SS (an SMBH binary) including all $\text{eight}$ planets ($\text{eight}$ stars)
with given positions $r_i$  and velocities $v_i$ measured
in units of orbit radius $r_p$ and velocity $v_p$ of Jupiter for SS
at $u/v_p=17$ (and of, e.g., the fifth
star for an SMBH binary at $u/v_p = 0.035$).
Thus in (\ref{eq2}) we have now for the SS $\text{eight}$ kick terms
with $J_i \sim (m_i/M) (v_i/v_p)^2$. For the SMBH binary model
we consider $\text{eight}$ stars modeled by map (\ref{eq2})
with the values  $J_1=2.5 \times10^{-4}$,
$J_2=5 \times 10^{-4}$, $J_3=7.5 \times10^{-4}$,
$J_4=10^{-3}$, $J_5=2.5 \times 10^{-3}$,  $J_6=6.25 \times 10^{-4}$,
  $J_7=5 \times 10^{-4}$, and $J_8=1.25\times10^{-4}$
with the same ratio $r_i/r_p$ as for the SS. In both cases we injected
$N_J = 2.8 \times 10^{10}$ particles considering evolution
during $\tau_S$ orbital periods of Jupiter (fifth star). The steady-state
density distribution
is shown in Fig.~\ref{fig9}. For the SS, $\rho(r)$  is very close to the case
of only one Jupiter discussed above. This result is natural since
its mass is dominant in the SS. For the SMBH binary model we also find
a similar  distribution (see Fig.~\ref{fig4})
with a slightly slower decay of $\rho_v(r)$ with $r$
($\beta =2.06 \pm 0.002$) due to the contribution of more stars.
We obtain $\zeta=3000$ (SS) and $\zeta_g=3 \times 10^4$ (SMBH).
These two examples show that the binary model captures
the main physical effects of the DMP capture and evolution.

\section{Discussion}
Our results show that DMP capture and dynamics inside two-body and few-body
systems can be efficiently described by symplectic maps. 
The numerical simulations and
analytical analysis show that in the center of these systems the DMP volume density 
can be enhanced by a factor $\zeta_g \sim 10^4$ compared to its galactic value.
The values of $\zeta_g$ are highest for a high velocity $v_p $ of a planet or star rotating 
around the  system center.
%%change 
We note that our approach based on a symplectic map description
of the restricted three-body problem is rather generic.
Thus it can also be used to analyze comet dynamics, 
cosmic dust, and free-floating constituents of the Galaxy.

%-------------------------------------------------------------------

\bibliographystyle{aa} % style aa.bst
\bibliography{biblio}

\end{document}